\documentstyle[12pt]{article}

\newcommand{\be}{\begin{equation}}
\newcommand{\ee}{\end{equation}}
\newcommand{\bea}{\begin{eqnarray}}
\newcommand{\eea}{\end{eqnarray}}
\newcommand{\nn}{\nonumber\\}

\def\nbd{neighbourhood}
 
\def\cinf{C^\infty}
\def\half{\textstyle{1\over2}}  
\def\sgn{{\mathrm sgn}} 
\def\tr{{\mathrm tr}}
\def\det{{\mathrm det}}
\def\D{{\cal D}}
\def\k{{e_3}}
\def\i{{\mathrm i}}

\begin{document}

\vspace*{0.5cm}
\begin{center}
{\large \bf Stability analysis of 
some integrable Euler equations for $SO(n)$}
\end{center}

\vspace{1.2cm}

\begin{center}
L. Feh\'er${}^1$ and I. Marshall${}^2$ \\

\bigskip

{\em 
${}^1$Department of Theoretical Physics,
 University of Szeged \\
Tisza Lajos krt 84-86, H-6720 Szeged, Hungary \\
E-mail: lfeher@sol.cc.u-szeged.hu\\

\bigskip

${}^2$
Department of Mathematics, EPFL \\
1015 Lausanne, Switzerland\\
E-mail: ian.marshall@epfl.ch
}
\end{center}

\vspace{2.2cm}

\begin{abstract}
A family of special cases of the integrable Euler equations on $so(n)$ 
introduced by Manakov in 1976 is considered.
The equilibrium points are found and their stability is studied. 
Heteroclinic orbits are constructed that connect unstable equilibria 
and are given by the orbits of certain $1$-parameter subgroups of $SO(n)$. 
The results are complete in the case $n=4$ and incomplete for $n>4$. 
\end{abstract}

\newpage
\section{Introduction}
\setcounter{equation}{0}

Suppose that we have a Hamiltonian vector field defined on a 
Poisson space. A natural problem is to find its equilibrium 
points and to check whether or not they are stable. Stability is
 important in mathematical modelling, as it gives an indication 
which behaviour exhibited by a (mathematical) modelling system is
a reliable representative of behaviour in the corresponding (real) 
modelled system. In the present paper we adopt the following

\smallskip
\noindent 
{\bf Definition.}
Let ${\mathbf  X}_H$ be the Hamiltonian vector field corresponding to the 
function $H$. The  equilibrium  point $x$ of ${\mathbf  X}_H$ is 
{\it stable} if for any {\nbd} $U$ of $x$, there exists a {\nbd} $V$ of $x$ such that 
$\phi(0)\in V$ and 
$\dot\phi(t)={\mathbf  X}_H(\phi(t))$ implies 
that $\phi(t)\in U\ \forall t\geq 0$; 
otherwise $x$ is an {\it unstable} equilibrium of ${\mathbf  X}_H$.

The aim of the present work is to perform  
a stability analysis for certain integrable Euler equations
associated with the group $SO(n)$, focusing mainly on the case $n=4$, and in
addition to examine the heteroclinic orbits.
These equations represent a particularly simple
special case of the integrable Hamiltonian systems introduced 
by Manakov in \cite{Man}.
In the $n=3$ case they reduce to the classical Euler 
equations for the angular momentum 
of a free rigid body in the moving frame.
As explained in several mechanics textbooks,  
the qualitative behaviour of the 
solutions of the classical Euler equations is easily visualised 
in terms of their phase portrait, 
see for instance the picture on the cover of the book \cite{MR}.
This system lives on a coadjoint orbit  $S^2$ of $SO(3)$,
and it has the interesting feature that the unstable equilibria are 
connected by heteroclinic orbits that are given by great circles on 
the sphere $S^2$. 
Let us recall that a heteroclinic orbit in general consists of the points 
of a nontrivial  integral curve of a dynamical system 
and equilibrium points.
Since the great circles on $S^2$ are the orbits
of the 1-parameter subgroups of $SO(3)$, 
we shall in the $SO(n)$ case enquire about the existence of 
heteroclinic orbits that are orbits of 1-parameter subgroups of $SO(n)$;
we consider this to give rise to the most interesting results of the  paper.

This work is intended as a step towards a stability analysis of the 
full set of the Manakov systems \cite{Man} which, in 
addition to the special case
studied here, contains for example 
the $n$-dimensional rigid body of \cite{Mis},  
and has many interesting Lie-algebraic 
generalizations \cite{MisFom, FT}. 
In fact, most of our results 
are not difficult to extend. The result for which generalization presents problems
is the one described in Section 3.

For convenient reference later,  
we now recall some standard facts about the linearisation 
of a Hamiltonian dynamical system and its use in the stability 
analysis of the original system. 
Let $M$ be a Poisson space. Let $H\in\cinf(M)$ and suppose that $x$ is an 
equilibrium point of the Hamiltonian vector field ${\mathbf  X}_H$.
The linearisation at $x$ is a flow in $T_xM$ given by
\be
\dot v=({\cal L}_V{\mathbf  X}_H)(x),
\label{1.1}\ee
with $V$ any vector field such that $V(x)=v$.
By choosing any
system of local coordinates in a {\nbd} of $x$, this becomes a system of the 
form
$\dot v={\mathbf  L}v$, with $\mathbf  L$ a square matrix of the same size as 
the 
dimension of $M$. For a Hamiltonian system the eigenvalues of the linearisation
at $x$ come in groups of four, in the sense that if $\lambda$ is an 
eigenvalue of $\mathbf  L$ then $-\lambda$ is an eigenvalue of $\mathbf  L$ 
and so is the complex
conjugate $\bar\lambda$. 
The following statements are well known, see \cite{HS}.

\noindent
1. $x$ is unstable if the linearisation at $x$ of the 
system $\dot\phi={\mathbf  X}_H(\phi)$ 
has an eigenvalue with a positive real part. 
If no eigenvalues of the linearisation have positive real part, then all
eigenvalues have to be imaginary; in this case $x$ may or may not be stable.

\noindent 
2. $x$ is stable if there exists $f\in \cinf(M)$ 
for which $\{f,H\}=0$ in a {\nbd} of $x$ and 
\be
(i)\ df(x)=0,\qquad
\quad (ii)\ d^2 f(x) \hbox{ is definite.}
\label{*}\ee

If the rank of the Poisson bracket
is constant in some {\nbd} of $x$,  
then it is sufficient  that properties $(i)$
and $(ii)$ in (\ref{*}) be satisfied with respect to vectors tangent to the symplectic
leaf through $x$. 
This is discussed for example in \cite{Arnold}.
In this paper we consider such a ``regular situation'' since the phase spaces
of our interest will be generic coadjoint orbits of the Lie group $SO(n)$.
We will assume all entities appearing in 
the definition of the systems studied to be generic, since 
this would be a reasonable assumption in a physical context 
and it also simplifies the problem. 

The organization of the paper and of our results is as follows.
The next section contains the definition of the Hamiltonian systems 
of  interest associated with $SO(n)$ 
together with a description of their equilibrium points (Proposition 1).
In section 3 we present a complete analysis of the stability of
the equilibrium points in the $n=4$ case. 
The outcome of our study is given by Proposition 2.
In section 4 we describe a necessary condition (Proposition 3) 
for the possibility 
to construct heteroclinic orbits by means of $1$-parameter subgroups
for Hamiltonian systems living on a coadjoint orbit, and 
concretely construct such heteroclinic orbits for the 
systems associated with $SO(4)$.
In section 5 the main features of the stability analysis are outlined for any $n$.
In particular, the construction of the heteroclinic orbits is generalized to the
$SO(n)$ case (see Proposition 4).
Section 6 contains a brief summary of the results and some open problems.

\section{A family of integrable Euler equations for $SO(n)$}
\setcounter{equation}{0}

We define below the Hamiltonian systems to be studied    
and describe their equilibrium points.
As explained at the end of the section, these systems correspond to 
a special case of the integrable Euler equations introduced in \cite{Man}.  
 
Consider  the Lie algebra $so(n)$ of the real orthogonal group $SO(n)$.
An element of $so(n)$ is an $n\times n$ antisymmetric real matrix.
The Lie-Poisson bracket of functions on $so(n)^*$ is given by
\be
\{\phi,\psi\}(\alpha)=<\alpha,[d_\alpha\phi,d_\alpha\psi]>
\qquad \forall \alpha\in so(n)^*, 
\label{2.1}\ee
where $d_\alpha\phi\in so(n)$ is defined by
\be
<\beta,d_\alpha\phi>=\left.{d\over dt}\right\vert_{t=0}\phi(\alpha+t\beta)
\qquad\forall\beta\in so(n)^*,
\label{2.2}
\ee
and $d_\alpha\psi$ similarly. The symplectic
leaves in $so(n)^*$ are the coadjoint orbits of $SO(n)$ in $so(n)^*$. 
It will be convenient to identify $so(n)^*$ with $so(n)$ 
with the aid of a multiple of the standard
trace form for $n\times n$ matrices, so 
that $<\beta,X>:=-\frac{1}{2}{\rm tr}\,(\beta X)$.

Let us define the Cartan subalgebra 
$\mathbf  h$ in $so(n)$ to be the set of all matrices $x$ 
of the form
\be
x=\sum_{k=1}^m x_k e_{kk} \otimes \i\sigma_2
\quad \hbox{if $n=2m$},
\quad\hbox{or}\quad 
x=\left({\matrix{\sum_{k=1}^m x_k e_{kk} \otimes \i\sigma_2&0\cr 0&0}}\right)
\quad \hbox{if $n=2m+1$},
\label{2.3*}\ee
where $m$ is any positive integer.
Here $e_{ij}$ is the $m\times m$  matrix having 1 for the term in the 
$i$th row and in the $j$th column and all other terms zero.
We use the Pauli matrices 
\be
\sigma_1=\left({\matrix{0&1\cr1&0}}\right),
\quad
\sigma_2=\left({\matrix{0&-\i\cr \i&0}}\right),
\quad
\sigma_3=\left({\matrix{1&0\cr0&-1}}\right),
\quad
\sigma_0:= {\mathbf  1}=\left({\matrix{1&0\cr0&1}}\right).
\label{Pauli}\ee

An element $x$ of $\mathbf  h$ is generic if $x_k\neq0\ \forall k$ 
and $x_k^2\neq x_l^2$ if $k\neq l$.
Using the identification of $so(n)^*$ with $so(n)$,
a generic symplectic leaf can be written as 
\be
{\cal O}_x=\{gxg^{-1}\vert g\in SO(n)\}
\label{2.4}\ee
with $x$ a generic element in $\mathbf  h$. 
The isotropy subalgebra of $x$ in $so(n)$ consists of 
matrices of the same form as $x$, i.e. given by the same formula as (2.3) 
with different values of $x_i$. The isotropy subgroup $SO(n)_x$ is the exponential of 
this algebra.

In this paper we are interested in 
Hamiltonian systems $({\cal O}_x,\{\ ,\ \}, H)$ on generic coadjoint orbits, 
where $H$ has the form 
\be
H(\mu):=-\frac{1}{2}{\rm tr}\,(J\mu^2),
\qquad
\mu \in {\cal O}_x, 
\label{2.5}\ee
with some constant matrix $J={\mathrm{diag}}(J_1,\dots,J_n)$. 
We assume that $J_i^2\neq J_j^2$ if $i\neq j$.
The generalized Euler equation defined by 
the Hamiltonian vector field ${\mathbf  X}_H$ can be written as follows:
\be
\dot\mu=[J,\mu^2].
\label{2.6}\ee

An equilibrium point on ${\cal O}_x$, for $x$ given by (2.3), is a point $gxg^{-1}$
such that
\be
0=[J,gx^2g^{-1}].
\label{2.7}\ee
Let $p$ be an element of the permutation group $S_n$ (the Weyl group of $sl(n)$),
and introduce the permutation matrix  $\bar p\in O(n)$ by
\be
\bar p_{ij} = \delta_{i, p(j)}\qquad (i,j=1,\ldots, n).
\ee
For any diagonal matrix $D={\mathrm diag}(d_1,\ldots, d_n)$, one has
\be
p(D):= {\mathrm diag}(d_{p^{-1}(1)},\ldots, d_{p^{-1}(n)})= \bar p D \bar p^{-1},
\ee
and the parity of $p$ satisfies $\sgn(p )=\det(\bar p)$.
Since  $J$ and $x^2$ are diagonal matrices,  
$\bar p x\bar p^{-1}$ is clearly an equilibrium point whenever it belongs 
to  ${\cal O}_x$. 
This holds obviously for the even permutations.
If $n=(2m+1)$ is odd, then $\bar p x\bar p^{-1}\in {\cal O}_x$ for any $p\in S_n$,
since in this case 
\be
\bar p D \bar p^{-1} = \hat p D \hat p^{-1}
\quad
\hbox{with}\quad
\hat p:= \sgn(p) \bar p \in SO(2m+1).
\ee 
We can prove that the equilibrium points associated in this manner with
the permutations exhaust all the equilibria on ${\cal O}_x$.

\medskip
\noindent
{\bf Proposition 1.}
{\em 
The set of equilibrium points on a generic orbit ${\cal O}_x$, for $x$ of the 
form given in (2.3), consists of the matrices $\bar p x\bar p^{-1}$,
where $p\in S_n$ is an even permutation if $n$ is even, and $p\in S_n$ is an 
arbitrary permutation if $n$ is odd.
The equilibrium points associated with different permutations are different. 
}
\medskip

\noindent
{\em Proof.}  Let us consider the set 
\be
E_{x}:= \{ gx g^{-1} \vert [J, gx^2 g^{-1}]=0,\quad g\in O(n)\}.
\label{P1}\ee
Since $J$ is a regular diagonal matrix by assumption, $g x^2 g^{-1}$ 
must be a diagonal matrix whose entries are obtained by permuting 
the entries of the diagonal matrix $x^2$. 
We  can choose a set of elements of $S_n$,
say  $\{ p_i\}_{i=1}^N$,  
for which  the matrices $p_i(x^2)$ are distinct from each other
for $i\neq j$ and they contain all matrices that are obtained
by permuting the diagonal entries of $x^2$.
Note that $N=\frac{n!}{2^m}$ for $n=2m$ or $n=(2m+1)$,
and the $p_i$ are a set of representatives for the 
coset space $S_n/S_n^{x^2}$, where 
\be
S_n^{x^2}=\{ p\in S_n \vert p(x^2)=x^2\}.
\ee
For $n=2m$ or $n=(2m+1)$,
the group $S_n^{x^2}$ is generated by the elements
\be
\tau^{1,2},
\tau^{3,4},
\ldots, \tau^{2m-1,2m},
\ee
where $\tau^{k,l}\in S_n$ denotes the transposition that exchanges $k$ with $l$. 

Since any $g\in O(n)$  that appears 
in (\ref{P1}) satisfies $g x^2 g^{-1}= \bar p_i x^2 \bar p_i^{-1}$ with some 
$1\leq i\leq N$, it follows that the most general such $g$ can be written as 
\be
g= \bar p_i \gamma 
\quad\hbox{with some}\quad
\gamma\in O(n)_{x^2},
\label{P3}\ee
\be
O(n)_{x^2}:=\{ \gamma\in O(n)\vert \gamma x^2 \gamma^{-1} = x^2\}.
\label{P4}\ee
The isotropy group $O(n)_{x^2}$
consists of block-diagonal matrices with arbitrary elements of 
$O(2)$ in the $2\times 2$ blocks.   
It is useful to consider also 
\be
O(n)_x:= \{ q\in O(n)\vert q x q^{-1}=x\},
\label{P5}\ee
which consists of block-diagonal
matrices with each $2\times 2$ block containing  an arbitrary element of $SO(2)$.
The point to notice is that any $\gamma\in O(n)_{x^2}$ can be 
uniquely written in the form\footnote{In fact, $O(n)_x$ is a normal subgroup of
$O(n)_{x^2}$ and $S_n^{x^2}$ is the corresponding factor group.}
\be
\gamma = \bar \Gamma q,  
\qquad \Gamma \in S_n^{x^2},\,q\in O(n)_x.
\label{P6}\ee
This follows
from the fact that $O(2)/SO(2)$ can be identified with 
the group generated by the transposition matrix $\sigma_1$. 
By using these observations, we see that any element 
$g x g^{-1}\in E_x$ has the form 
\be
g x g^{-1}= \bar p_i \bar \Gamma  x  (\bar p_i \bar \Gamma )^{-1}
\qquad
(1\leq i \leq N,\,\, \Gamma\in S_n^{x^2}).
\label{P7}\ee
As $\bar p= \bar p_i  \bar \Gamma$ for $p= p_i  \Gamma$,
this implies that all elements of $E_x$ are given by
$\bar p x \bar p^{-1}$ with some $p\in S_n$.
It is clear from the definitions 
that any $p\in S_n$ can be decomposed as $p=p_i \Gamma$ with a unique 
$p_i$ and a unique element of $S_n^{x^2}$,
and one can check directly that  different permutations 
are associated with different points of $E_x$. 

If $n=(2m+1)$, the statement of the proposition follows 
immediately  from the above-established  results 
(the coadjoint orbits of $O(2m+1)$ and $SO(2m+1)$ coincide). 
The proof is completed by noting that
in the $n=2m$ case only those
elements  $\bar p x \bar p^{-1}$
lie on the coadjoint orbit of $SO(2m)$ through $x$ for which $p$
is an even permutation.
This fact can be verified, for example, by performing an analogous
analysis as above  in the case for which $g$ in (2.12) is restricted
to $SO(n)$ from the beginning.

\medskip

\noindent
{\bf Remark 1.}
Suppose that we study the nature of an equilibrium point 
$\bar p x\bar p^{-1}\in {\cal O}_x$. 
We may then choose a different basis in which this point is represented 
by the matrix of $x$, and  $J$ is replaced by the matrix $\bar p^{-1}J\bar p$.
We may thus assume without loss of generality that the equilibrium point of interest
is always represented by the same matrix $x\in{\mathbf  h}$ in (2.3).

\noindent
{\bf Remark 2.}
It follows from (\ref{2.6}) that in our case 
the linearised system at the point $x$ is the flow in $T_x{\cal O}_x$ defined by 
\be
\dot{v}=[J,v x + x v].
\label{2.8}\ee 

\noindent
{\bf Remark 3.}
In the terminology of generalized rigid bodies \cite{Arnold} 
the quantity $\mu$ in the Euler
equation (\ref{2.6}) is the angular momentum relative to the body.
Correspondingly, the inverse of the moment of inertia 
operator maps $\mu$ to the angular velocity $\omega$ relative to the body 
according to  $\mu \mapsto \omega= - (J \mu + \mu J)$.  
Indeed, then (\ref{2.6}) takes the classical form 
$\dot{\mu}=[\mu, \omega]$.
This is a special case of the integrable rigid body systems introduced in \cite{Man}
by the relation $\mu_{ij} = \frac{a_i - a_j}{b_i - b_j} \omega_{ij}$
with  arbitrary constants $a_i$, $b_i$. 
The case (\ref{2.6}) arises by setting $b_i=a_i^2$ with $a_i=-J_i$,
while the $n$-dimensional
rigid body of \cite{Mis} is obtained by setting $a_i=b_i^2$.    

\section{Stability analysis in the $n=4$ case}
\setcounter{equation}{0}

The Lie algebra $so(4)$ is the same as the direct sum $so(3)\oplus so(3)$. 
This can be seen by identifying
$so(3)$ with $su(2)$ and then finding two commuting copies of $su(2)$ in $so(4)$.
In terms of the Pauli matrices,
we have $su(2)=span\{\i\sigma_1, \i\sigma_2, \i\sigma_3\}$, and 
two commuting $su(2)$ subalgebras that together span $so(4)$ are 
\be
su(2)\cong 
{\mathrm{span}}\{\sigma_1\otimes \i\sigma_2, \i\sigma_2\otimes \sigma_0, 
\sigma_3\otimes \i\sigma_2\}
\cong 
{\mathrm{span}}\{i\sigma_2\otimes\sigma_1, 
\sigma_0\otimes \i\sigma_2, \i\sigma_2\otimes\sigma_3\}.
\label{3.1}\ee
In the coordinates $(l,m)$ on $so(4)=so(3)\oplus so(3)$ given by 
\be
\mu=
\bigl[
l_1\sigma_1\otimes \i\sigma_2 + l_2\i\sigma_2\otimes \sigma_0 + 
l_3\sigma_3\otimes \i\sigma_2
\bigr]
-\bigl[
m_1\i\sigma_2\otimes\sigma_1+m_2\i\sigma_2\otimes\sigma_3 + 
m_3\sigma_0\otimes \i\sigma_2
\bigr],
\label{3.2}\ee
the Poisson bracket is
\be
\{l_i,l_j\}=\epsilon_{ijk}l_k,\qquad\{l_i,m_j\}=0,
\qquad\{m_i,m_j\}=\epsilon_{ijk}m_k,
\label{3.3}\ee
and  $\vert l\vert^2$ and $\vert m\vert^2$ are Casimir functions. 
The rigid body Hamiltonian $H$ and an independent commuting integral $K$ are 
now given by
\be
H(l,m)=l^T\Lambda m,
\qquad
K(l,m)=\half l^T\Lambda^2 l+\half m^T\Lambda^2 m 
- l^T\Theta m,
\label{3.4}\ee
where $\Lambda$ and $\Theta$ are constant diagonal matrices  
\be
\Lambda = {\mathrm{diag}}(\Lambda_1, \Lambda_2, \Lambda_3),
\qquad
\Theta= {\mathrm{diag}}(\Lambda_2 \Lambda_3, \Lambda_1 \Lambda_3, \Lambda_1 \Lambda_2).
\ee 
The equations of motion corresponding to $H$ are
\be
\dot l=(\Lambda m)\wedge l\qquad \dot m=(\Lambda l)\wedge m.
\label{3.5}\ee
We apply the usual identification of $so(3)$ with ${\mathbf  R}^3$ equipped
with the vector-product, denoted by $\wedge$.
The formulae in (\ref{3.4})  may be recovered by a technique due to
Manakov \cite{Man}: define ${\cal L}=\lambda J+\mu$, then the set of coefficients of 
$\lambda$ amongst all traces of powers of ${\cal L}$ forms a commuting family. 
Here $K\sim(\tr {\cal L}^4)\vert_{\lambda^2}$, $H\sim(\tr {\cal L}^3)\vert_{\lambda^1}$
up to Casimirs and (from (\ref{2.5}))
$\Lambda$ is related to $J$ by 
$\Lambda_1=-J_1+J_2+J_3-J_4$, $\Lambda_2=-J_1+J_2-J_3+J_4$,
$\Lambda_3=-J_1-J_2+J_3+J_4$.  
{}From now on we make the {\em genericity assumption} that
\be
\Lambda_i\!{}^2 \neq \Lambda_j\!{}^2
\quad\hbox{if}\quad i\neq j
\quad\hbox{and}\quad \Lambda_i\neq 0
\quad\forall i.
\label{genLambda}\ee
The first part of these conditions follows from the 
assumption that $J_p^2 \neq J_q^2$ for $p\neq q$. 

Let $e_k$ ($k=1,2,3$) denote the standard basis of ${\mathbf  R}^3$. 
The equilibrium points of (\ref{3.5}) 
that lie on generic coadjoint orbits are in fact given by   
\be
(l,m)=b (a e_k, e_k)
\quad\hbox{with }\quad {\mathbf  R}\owns a,b\neq 0,
\quad
k=1,2,3.
\label{genzero}\ee
We next study the stability of an equilibrium point of the form  
\be
(l,m)=(a\k,\k) \quad\hbox{ with }\quad {\k}=(0,0,1)^T 
\quad \hbox{and}\quad 
{\mathbf  R}\owns a\neq0,
\label{3.6}\ee
and then the general case (\ref{genzero}) will be reduced 
to this one. 

The elements of $T_{(ae_3, e_3)} {\cal O}_{(ae_3,e_3)}$
can be parametrized as $(a \xi\wedge e_3, \eta \wedge e_3)$ with 
$\xi= \xi_1 e_1 + \xi_2 e_2$ and $\eta= \eta_1 e_1 + \eta_2 e_2$.
By putting $v=(\xi_1, \eta_1, \xi_2, \eta_2)^T$,   
the linearised system at $(a{\k},{\k})$ is given 
explicitly by $\dot v=Lv$ with
\be
L=\left({\matrix{0&0&-\Lambda_3&\Lambda_1\cr0&0&a\Lambda_1&-a\Lambda_3\cr
\Lambda_3&-\Lambda_2&0&0\cr-a\Lambda_2&a\Lambda_3&0&0}}\right).
\label{3.7}\ee
The eigenvalues $\zeta$ of $L$ satisfy
\be
\zeta^4+\Bigl[(a^2+1)\Lambda_3\!{}^2+2a\Lambda_1\Lambda_2\Bigr]\zeta^2 
+ a^2(\Lambda_3\!{}^2-\Lambda_1\!{}^2)(\Lambda_3\!{}^2-
\Lambda_2\!{}^2)=\det(\zeta I-L)=0.
\label{3.8}\ee
Let $\D$ be defined by 
\begin{eqnarray}
&&\D:=\Bigl[
(a^2+1)\Lambda_3\!{}^2+2a\Lambda_1\Lambda_2\Bigr]^2
-4a^2(\Lambda_3\!{}^2-\Lambda_1\!{}^2)(\Lambda_3\!{}^2-\Lambda_2\!{}^2)\nn
&&\phantom{\D:}=\Lambda_3\!{}^2\Bigr[(a^2-1)^2\Lambda_3\!{}^2+
4a^2(\Lambda_1\!{}^2+\Lambda_2\!{}^2)
+4a(a^2+1)\Lambda_1\Lambda_2\Bigr].
\label{3.9}\end{eqnarray}
Stability of the equilibrium point $(a{\k},{\k})$ requires all roots of (3.8) to
be imaginary. Hence all three of the following conditions must be fulfilled:
\bea
&&(i)\qquad(\Lambda_3\!{}^2-\Lambda_1\!{}^2)(\Lambda_3\!{}^2-\Lambda_2\!{}^2)>0,\nn
&&(ii)\qquad(a^2+1)\Lambda_3\!{}^2+2a\Lambda_1\Lambda_2>0,\label{**}\\
&&(iii)\quad a)\ \D>0,\quad\hbox{ or }\quad b)\ \D=0.
\nonumber\eea
If any one of the conditions of (\ref{**}) is not satisfied 
then (\ref{3.8}) has roots of the 
form $\zeta=\pm\alpha\pm i\beta$, with $\alpha\neq0$, 
and the equilibrium point (\ref{3.6}) is unstable.

Suppose that $\D=0$. Every neighbourhood of $(a{\k},{\k})$ contains points 
of the form $\bigl((a\pm\epsilon){\k},{\k})$ with $\epsilon>0$. As $\D<0$ at 
one of these two points, it follows that there are unstable equilibrium points
arbitrarily close to $(a{\k},{\k})$ and hence $(a{\k},{\k})$ is unstable.
We have instability  then if (i), (ii), (iiib) of (\ref{**}) are 
satisfied despite all the 
eigenvalues of the linearised system being pure imaginary.

We shall prove stability in the case (i), (ii), (iiia) of (\ref{**}) by
exhibiting a constant of motion for which (\ref{*}) holds.
As a preparation let us introduce  
\be
F:=\Lambda_1\Lambda_2H+\Lambda_3K 
\label{3.10}
\ee
and denote by $\tilde H$ and $\tilde F$ the restrictions of $H$ and $F$ to 
the orbit through the equilibrium point (\ref{3.6}).
One can check that $d \tilde F=0$ at  $(a{\k},{\k})$ and, up to a
common constant of proportionality, the Hessians of $\tilde H$ and $\tilde F$ 
at this critical point are found to be
\be
d^2\tilde H\sim\left({\matrix{{\mathbf  H}_1&\mathbf {0}\cr{\mathbf  0}&{\mathbf  H}_2}}\right)
\label{3.11}\ee
and 
\be
d^2 \tilde F\sim\left({\matrix{{\mathbf  F}_1&\mathbf {0}\cr{\mathbf  0}&{\mathbf  F}_2}}\right)
\label{3.12}\ee
where ${\mathbf  H}_1,{\mathbf  H}_2,{\mathbf  F}_1,{\mathbf  F}_2$ are the following $2\times2$ matrices:
\be
{\mathbf  H}_1=\left({\matrix{\Lambda_3-\Lambda_2&0\cr0&\Lambda_3+\Lambda_2}}\right),
\qquad
{\mathbf  H}_2=\left({\matrix{\Lambda_3-\Lambda_1&0\cr0&\Lambda_3+\Lambda_1}}\right),
\ee
\be
{\mathbf  F}_1=(\Lambda_3\!{}^2-\Lambda_2\!{}^2)
\left({\matrix{a+1&a-1\cr a-1&a+1}}\right)
\left({\matrix{\Lambda_3+\Lambda_1&0\cr0&\Lambda_3-\Lambda_1}}\right)
\left({\matrix{a+1&a-1\cr a-1&a+1}}\right),
\ee
\be
{\mathbf  F}_2=(\Lambda_3\!{}^2-\Lambda_1\!{}^2)
\left({\matrix{a+1&a-1\cr a-1&a+1}}\right)
\left({\matrix{\Lambda_3+\Lambda_2&0\cr0&\Lambda_3-\Lambda_2}}\right)
\left({\matrix{a+1&a-1\cr a-1&a+1}}\right).
\ee

\medskip
\noindent
{\bf Lemma.}
{\em If (i), (ii), (iiia) of (\ref{**}) are all satisfied then 
the equilibrium point (\ref{3.6}) is stable.}
\medskip

\noindent
{\em Proof.} There are two cases to consider.

\noindent
{\em Case one:} 
$\Lambda_3\!{}^2-\Lambda_1\!{}^2>0$ and $\Lambda_3\!{}^2-\Lambda_2\!{}^2>0$. 
In this case (i) clearly holds. It is obvious that $d^2\tilde H$ is either positive or 
negative definite at $(a\k,\k)$ and the same applies to $d^2\tilde F$. Of course
it can be shown that (ii) and (iiia) also hold. 
 
\noindent
{\em Case two:}
$\Lambda_3\!{}^2-\Lambda_1\!{}^2<0$ and $\Lambda_3\!{}^2-\Lambda_2\!{}^2<0$. 
In this case again (i) clearly holds. Let us additionally suppose that (ii) and (iiia)
both hold. We can show that there exists a $z\in{\mathbf  R}$ such that 
$d^2(4z \tilde H+\tilde F)$ is definite at $(a{\k},{\k})$.

The details of the proof in case two are as follows. Let us write
\be
d^2(4z \tilde H+\tilde F)\sim
\left({\matrix{{\mathbf  Q}_1&0\cr0&{\mathbf  Q}_2}}\right):={\mathbf  Q},
\label{3.14}
\ee 
with the $2\times2$ matrices ${\mathbf  Q}_i=4z {\mathbf  H}_i+{\mathbf  F}_i$.
Now $\mathbf  Q$ is a
positive or negative definite matrix if and only if 
\be
\det{\mathbf  Q}_1>0,\quad
\det{\mathbf  Q}_2>0,\quad
\,\,\,\hbox{and }\,\,\, 
\tr{\mathbf  Q}_1 \tr{\mathbf  Q}_2>0.
\label{3.15}\ee
The first and second conditions of (\ref{3.15}) require
\be
z^2+\Bigl[(a^2+1)\Lambda_3\!{}^2+2a\Lambda_1\Lambda_2\Bigr]z
+ a^2(\Lambda_3\!{}^2-\Lambda_1\!{}^2)(\Lambda_3\!{}^2-\Lambda_2\!{}^2)<0.
\label{3.16}\ee
Notice that (\ref{3.16}) is similar to (\ref{3.8}). 
Now (i), (ii), (iiia) together are equivalent to (\ref{3.8}) having  
four distinct, imaginary eigenvalues, and this is obviously equivalent 
to the solvability of  (\ref{3.16}) for $z\in{\mathbf  R}$. 
Let us write a solution $z$ in the form 
\be
z=-\half\Bigl[(a^2+1)\Lambda_3\!{}^2+2a\Lambda_1\Lambda_2\Bigr]+\half\beta.
\label{3.17}\ee
Then (\ref{3.16}) implies 
\be
\beta^2<\D
\label{3.18}\ee
and because of (i),
\be
\beta <(a^2+1)\Lambda_3\!{}^2+2a\Lambda_1\Lambda_2.
\label{3.19}\ee
Using  (\ref{3.17}) we obtain 
\be
\tr{\mathbf  Q}_1 \tr{\mathbf  Q}_2
 =
\Bigl(8z\Lambda_3+4(\Lambda_3\!{}^2-\Lambda_2\!{}^2)\Lambda_3(a^2+1)\Bigr)
\Bigl(8z\Lambda_3+4(\Lambda_3\!{}^2-\Lambda_1\!{}^2)\Lambda_3(a^2+1)\Bigr)
=16\Lambda_3\!{}^2 X Y, 
\ee
with 
\bea
&&X=
\Bigl(\beta-\bigl[(a^2+1)\Lambda_3\!\!{}^2+2a\Lambda_1\Lambda_2\bigr]
+(a^2+1)(\Lambda_3\!\!{}^2-\Lambda_2\!\!{}^2\Bigr),\nn
&& Y= 
\Bigl(\beta-\bigl[(a^2+1)\Lambda_3\!\!{}^2+2a\Lambda_1\Lambda_2\bigr]
+(a^2+1)(\Lambda_3\!\!{}^2-\Lambda_1\!\!{}^2\Bigr).
\label{3.20}\eea
Eq.~(\ref{3.19}) with the assumptions 
$\Lambda_3\!{}^2-\Lambda_1\!{}^2<0$ and $\Lambda_3\!{}^2-\Lambda_2\!{}^2<0$
imply that $X<0$ and $Y<0$, and hence $\tr{\mathbf  Q}_1 \tr{\mathbf  Q}_2>0$.
Since all three conditions (\ref{3.15}) for the 
definiteness of $\mathbf  Q$ are satisfied,
$f:=(4z \tilde H+\tilde F)$ satisfies (\ref{*}) 
at the equilibrium point $(a{\k},{\k})$,
whereby the proof is complete.

\medskip
The results proven above imply the following 
proposition, which provides a characterization of the stability of the 
equilibrium points of (\ref{3.5}) on generic coadjoint orbits.

\medskip
\noindent
{\bf Proposition 2.}
{\em 
The equilibrium point $b (ae_k, e_k)$ in (\ref{genzero}) is stable 
if and only if (i), (ii), (iiia) of (\ref{**}) hold for the constant $a$ and 
the matrix $\Lambda$ replaced by the matrix 
$\Lambda^P:= diag (\Lambda_{P(1)}, \Lambda_{P(2)}, \Lambda_{P(3)})$
where $P$ is an even permutation of $(1,2,3)$ for which $P(3)=k$.}
\medskip

The permutation part of the statement follows obviously from (\ref{3.5}) 
after checking that the stability of $b(a e_3, e_3)$ is equivalent to 
the stability of $(a e_3, e_3)$. 
In general, equation (\ref{2.6}) has the property 
that $\mu(t)$ is a solution if and only if $\mu_b(t):= b \mu(bt)$
is a solution for any $b\neq 0$. 
This implies the required result for $b>0$.
The $b=-1$ case is settled by using the facts that the 
matrix of the linear system (\ref{2.8}) simply gets multiplied by $-1$ 
under such a rescaling of the equilibrium point, while 
the conserved quantities $H$ and $K$ in (\ref{3.4}), 
and thus also their second variations,  remain unchanged. 

\section{Heteroclinic orbits from 1-parameter subgroups}
\setcounter{equation}{0}

Consider two equilibrium points, $x_0$ and $x_1$, of a  
smooth Hamiltonian vector field  ${\mathbf  X}_H$ on a 
coadjoint orbit ${\cal O}$ of a compact Lie group $G$
with Lie algebra  $\mathbf  g$.
Let us look for a 1-parameter subgroup of $G$
that generates a heteroclinic orbit of  ${\mathbf  X}_H$ 
connecting these equilibria.
For $Y \in {\mathbf  g}$, define 
\be
\gamma(s)=Ad^*_{\exp(s Y)}x_0.
\label{4.1}\ee
Then our first requirement is that $\gamma(s_1)=x_1$ for some $s_1>0$. 
Setting $s_0:= 0$, our second requirement is that the curve
$\gamma: (s_0, s_1) \rightarrow {\cal O}$ yields an integral curve 
of ${\mathbf  X}_H$ by a suitable reparametrization.
In other words, there should exist an increasing  
diffeomorphism  
$T: (s_0, s_1) \rightarrow (-\infty, +\infty)$ for 
which the curve $c(t)$ defined by
\be
c(T(s))= \gamma(s) 
\qquad
\forall s\in (s_0, s_1)
\label{4.2}\ee
satisfies $\dot{c}(t) ={\mathbf  X}_H(c(t))$ for any  $t\in {\mathbf  R}$.   
Denoting the derivative with respect to $s$ by prime, it follows that 
$\forall  s\in (s_0, s_1)$ we have
 \be
\chi(s)\gamma'(s)={\mathbf  X}_H(\gamma(s))
\quad
\hbox{with}\quad
\chi(s) = \frac{1}{T'(s)}.
\label{4.3}\ee
Because of the smoothness of the right hand side 
as a function of $s\in {\mathbf  R}$, 
we observe that a unique extension of $\chi$ to $[s_0, s_1]$ must exist.
This extended function must clearly satisfy the conditions 
\be
\chi(s_0)=\chi(s_1)=0,
\qquad
\chi'(s_0 +0) \geq 0,
\quad
\chi'(s_1-0)\leq 0.
\label{4.4}\ee
By using that (\ref{4.3}) holds on $[s_0, s_1]$ and taking 
the appropriate derivatives of this equality at the endpoints, 
one arrives at the following statement.  

\medskip
\noindent
{\bf Proposition 3.}
{\em If $\gamma(s)$ in (\ref{4.1})  yields a  heteroclinic orbit in the
above-described sense, then  the vectors
$ad^*_Y x_i\in T_{x_i}{\cal O}$
are eigenvectors of the linearisation of ${\mathbf  X}_H$ at $x_i$, for $i=0,1$,  
with the respective eigenvalues being 
$\chi'(s_0 +0)$ and $\chi'(s_1-0)$. 
}
\medskip

In particular, notice from the proposition that the existence 
of a real eigenvalue of the linearisation of ${\mathbf  X}_H$ at $x_0$ is a
{\em necessary} condition for the construction of a heteroclinic orbit
through $x_0$ by means of a 1-parameter subgroup of $G$. 
For the rigid body systems described in section 2, 
this is in fact also a {\em sufficient} condition. 
For $n=3$ this is a well known result. 
We verify it below in the 
$n=4$ case by using the explicit analysis of the preceding section.

 As before we may assume that the
equilibrium point of interest is $x_0=(a{\k},{\k})$ in (\ref{3.6}). 
Let $(a\xi\wedge{\k},\eta\wedge{\k})= [Y, x_0]$ be an 
eigenvector of the linearised flow
at $x_0$ with real eigenvalue $z>0$.
Note that $z=0$ is excluded by (\ref{3.8}) and that we have $Y=(\xi,\eta)$
by using the identification of the Lie bracket of $so(3) \cong su(2)$
with the vector-product.
Then we can check that 
\be
\xi_1^2+\xi_2\!{}^2=\eta_1\!{}^2+\eta_2\!{}^2.
\label{4.5}\ee
This follows from the eigenvector equation $Lv=zv$ with $L$ in (\ref{3.7}) 
and $v= (\xi_1, \eta_1, \xi_2, \eta_2)^T$. 
We set $\Delta:= \sqrt{\xi_1^2 + \xi_2^2}$ and consider the curve  
\be
\gamma(s)= e^{s Y} x_0 e^{-sY}=
\cos(s\Delta)(a{\k},{\k}) + 
\Delta^{-1}\sin(s\Delta)(a\xi\wedge{\k},\eta\wedge{\k}).
\label{4.6}\ee
We can verify that this curve yields a heteroclinic orbit
that connects $x_0$ with $x_1:= - (a\k,\k)$ for
$s_0=0$ and $s_1=  \frac{\pi}{\Delta}$.
Indeed, the functions $\chi$ and $T$ introduced in (\ref{4.3}) are found as 
\be
\chi(s)=\frac{z}{\Delta}\sin(s\Delta),  
\label{4.7}\ee
\be
T(s) = \frac{1}{z} \log \tan \frac{s\Delta}{2}
\quad\hbox{for}\quad 
s_0 <s < s_1. 
\label{4.8}\ee 

Note that the adjoint and coadjoint actions are the same for any compact Lie 
group and any orbit ${\cal O}_x= G/G_x$
carries a canonical $G$-invariant Riemannian metric induced 
by the Cartan-Killing form on ${\mathbf  g}$.
It is well known that the geodesics of this metric 
coincide with the orbits of the 1-parameter subgroups of $G$.
Thus the heteroclinic orbits considered above 
are proper generalizations of the heteroclinic orbits of the standard 
rigid body that are great circles on $S^2= SO(3)/SO(2)$.
  
\section{On the stability analysis for $n>4$}
\setcounter{equation}{0}

We are able to repeat a large part of the stability analysis 
performed in the 4-dimensional case.
Specifically: we can find the equilibrium points (Proposition 1); we can 
find the eigenvectors  and corresponding eigenvalues of the linearised system at 
each equilibrium point; we can prove the converse 
of Proposition 3.
However the problem of proving 
stability (or not) for the equilibrium points having all 
eigenvalues pure imaginary is 
more complicated.
We  present here only an outline of the stability analysis for general $n$.

To find the eigenvalues and eigenvectors of the linearised 
system (\ref{2.8}) at $x$
it is useful to decompose $so(n)$ as the 
vector space direct sum $so(n)=$ Ker$(ad_x)+$ Im$(ad_x)$, whereby 
we can uniquely parametrize $v\in T_x {\cal O}_x$ as 
$v=[Y, x]$ with $Y\in$ Im$(ad_x)$. 
The linearised system (\ref{2.8}) then reads as 
\be
[\dot{Y},x]=[J,[Y,x^2]]
\label{5.1}\ee
and an eigenvector $[Y,x]\in T_x{\cal O}_x$ with eigenvalue $z$ satisfies
\be
[J,[Y,x^2]]=z [Y,x].
\label{5.2}\ee 
Let us take $x$ to be of the form (\ref{2.3*}) and
choose coordinates on $so(n)$ according to the natural decomposition into blocks.
That is, for $n=2m$ write 
$Y\in Im(ad_x)$  as $Y=A-A^T$ with 
$A=\sum_{i< j}e_{ij}\otimes\xi_{ij}$ and $\xi_{ij}$ a real $2\times2$ real matrix.
If $n=2m+1$, then $Im(ad_x) \owns Y=\left({\matrix{A-A^T&v\cr -v^T&0}}\right)$ with 
$A$ as before and $v^T=(v_1^T,v_2^T,\cdots,v_m^T)$ with $v_i$ a real $2\times1$ matrix.
Writing (\ref{5.2}) in these coordinates, we see directly that 
there are several copies of
the eigenvector equation for $so(4)$ - each of which has 4 solutions - and in the odd 
$n$ case also several copies of the eigenvector equation for $so(3)$ - each of which has
2 solutions. In fact we obtain exactly the right
number of such decoupled equations to generate all eigenvectors and their eigenvalues.
If any eigenvalue is real and nonzero, then we can use 
either the result described for $so(4)$ or a similar one - which has not been explicitly
described here, but which is straightforward - for $so(3)$, 
to construct  heteroclinic orbits  by suitable curves of the form in (\ref{4.1}). 
This leads to the following converse of Proposition 3. 

\medskip
\noindent 
{\bf Proposition 4.}
{\em 
Suppose that $z$ is a nonzero, real eigenvalue of the linear system (\ref{5.2}) at $x$.
Then there exists a corresponding eigenvector 
$[Y, x]$ for which the curve $\gamma(s)= e^{sY} x e^{-sY}$ yields a heteroclinic orbit
of the rigid body system (\ref{2.6}).} 
\medskip
 
We now sketch the proof of this proposition in the $n=2m$ case. 
In this case we can write $J= \sum_{i=1}^m e_{ii} \otimes D_i$, where the $D_i$
are  2 by 2 diagonal matrices. 
By putting 
\be
Y:= \sum_{1\leq i<j\leq m} Y_{ij}
\quad\hbox{with}\quad  
Y_{ij}:= e_{ij} \otimes \xi_{ij} - e_{ji} \otimes \xi_{ij}^T,
\label{Y1}\ee  
the eigenvector equation (\ref{5.2}) decouples into separate equations for each pair
of indices $i <j$,  
\be
(x_i^2 - x_j^2) ( D_i \xi_{ij} - \xi_{ij} D_j) = z (x_j \xi_{ij} S - x_i S \xi_{ij}),
\qquad
S:= \i \sigma_2.
\label{Y2}\ee
For any $1\leq i<j\leq m$, consider the $so(4)$ subalgebra of $so(2m)$ given by 
\be
so(4)_{ij} := {\mathrm{span}}\{e_{ii} \otimes S, e_{jj} \otimes S, 
( e_{ij} \otimes Q - e_{ji} \times Q^T)
\,\vert\, \forall Q\in gl(2, {\mathbf  R})\}.     
\label{Y3}\ee
The point to notice is that (\ref{Y2}) coincides with the eigenvalue equation 
for a rigid body system defined on $so(4)_{ij}$ at the corresponding equilibrium point
$x_{ij} := x_i e_{ii} \otimes S + x_j e_{jj} \otimes S$.
This implies by the $so(4)$ result established in section 4 that if 
$Y_{ij}$ is a solution of (\ref{Y2}) with some real $z\neq 0$, then the curve
\be
\gamma_{ij}(s) := e^{ s Y_{ij}} x_{ij} e^{-s Y_{ij}} 
\label{Y4}\ee
yields a heteroclinic orbit connecting the unstable equilibria $\pm x_{ij}$ 
of the induced rigid body system on $so(4)_{ij}$.
Decomposing $x$ as $x= x_{ij} + x'_{ij}$, we can check the relations 
\be
\gamma(s):= e^{ s Y_{ij}} x e^{-s Y_{ij}} = \gamma_{ij}(s) + x'_{ij}
\label{Y5}\ee
and 
\be
[ J, \gamma^2(s)] = [J_{ij}, \gamma_{ij}^2(s)],
\qquad
J_{ij}:= e_{ii} \otimes D_i + e_{jj}\otimes D_j.
\label{Y6}\ee 
Eq.~(\ref{Y6}) relates the Hamiltonian vector fields 
for the rigid  body systems on $so(2m)$ and on $so(4)_{ij}$ along the respective
curves $\gamma(s)$ and $\gamma_{ij}(s)$.
By collecting the above remarks, we conclude that $\gamma(s)$ in (\ref{Y5}) 
yields a heteroclinic orbit that connects the unstable equilibria 
$x'_{ij} \pm x_{ij}$.    

To illustrate what happens for odd $n$, let us look at $n=5$.
Let us assume that the equilibrium point $x$ of interest has the form
\be
x=\left({\matrix{
0&a-1&0&0&0\cr
-a+1&0&0&0&0\cr
0&0&0&-a-1&0\cr
0&0&a+1&0&0\cr
0&0&0&0&0}}\right),
\label{5.3}\ee
and parametrize $Y\in$ Im$(ad_x)$ according to 
\be
Y=\frac{1}{2}\left({\matrix{0&0&\xi_2-\eta_2&\xi_1-\eta_1&v_1\cr
		   0&0&-\xi_1-\eta_1&\xi_2+\eta_2&v_2\cr
		   \eta_2-\xi_2&\eta_1+\xi_1&0&0&w_1\cr
		   \eta_1-\xi_1&-\xi_2-\eta_2&0&0&w_2\cr
		   -v_1&-v_2&-w_1&-w_2&0}}\right).
\label{5.4}\ee
Then consider the eigenvector equation (\ref{5.2}) with 
$J= {\mathrm{diag}}(J_1, J_2, J_3, J_4, J_5)$.
By setting $v_i=0=w_i$ we reduce to the eigenvector condition for $so(4)$; by setting 
$\xi_i=\eta_i=w_i=0$ we reduce to the eigenvector condition for $so(3)$ and 
by setting $\xi_i=\eta_i=v_i=0$ we reduce to the eigenvector condition for $so(3)$ too.
In fact the coordinates have been chosen here so as to agree exactly with those used
for the so(4) analysis in section 3. In this way we find all 8 eigenvalues. 
The problems of checking if the eigenvalues are real, complex or imaginary reduce to
those of the $so(3)$ and $so(4)$ cases. Similarly  the 
construction of heteroclinic orbits as orbits 
of 1-parameter subgroups reduces to the $so(3)$ and $so(4)$ cases. To check if all 
eigenvalues of the linearisation being imaginary is sufficient for stability we
could try to prove the convexity at $x$ of a function of the form
\be
f=\alpha H + \beta H_1 +\gamma H_2 + \delta H_4
\label{5.7}\ee
with $H$, $H_1$, $H_2$, $H_3$ the Hamiltonian together with 3 independent commuting
integrals, which can be generated using the Lax 
matrix of Manakov \cite{Man}, where 
$\alpha$, $\beta$, $\gamma$, $\delta$ are expected to depend on the 
equilibrium point in question. Of course, while even this can be done in principle,
there is no strategy telling us how to proceed for general $n$.

\section{Conclusion}

In this paper we studied the equilibrium
points for the integrable Euler equations in  (\ref{2.6}).
In particular, we described the equilibrium points (Proposition 1) and
associated  heteroclinic 
orbits with any nonzero, real eigenvalue of the 
linearised system for any $n$ (Proposition 4).
We also found a complete characterization of the stability of the equilibrium points 
for $n=4$ (Proposition 2), but our     
stability analysis is incomplete for $n>4$.
In this case an open question is to find a criterion for the stability of those
equilibrium points for which all eigenvalues of the linearised system 
are imaginary.

As a final remark,  we wish to mention 
the work of Mishchenko and Fomenko \cite{MisFom} (for a review, see \cite{FT})
that contains generalizations of the systems 
of Manakov \cite{Man} to other Lie algebras. 
Various elements of our results have a general
Lie-algebraic nature and thus may be applicable to the systems of \cite{MisFom}. 
In this respect, it is natural to ask if Proposition 4 is valid 
only for the special cases (\ref{2.6}) that we considered here,
or can be extended to other systems among those  in \cite{Man, MisFom}, too.
It would also be interesting to find a general criterion of stability that could be used 
effectively to analyse these systems.

\section*{Acknowledgments}
We wish to thank T.S. Ratiu for posing the problem
studied here and for discussions.
L.F. was supported in part by the Hungarian 
Scientific Research Fund (OTKA) under T034170, T030099, T029802 and M036804.

\end{document}